\documentclass{article}
\usepackage{spconf,amsmath,graphicx}
\usepackage[colorlinks,linkcolor=blue]{hyperref}
\usepackage{caption}
\usepackage{multirow}
\usepackage{float}
\usepackage{setspace}
\usepackage{booktabs}
\ninept

\usepackage{amsmath}
\usepackage[psamsfonts]{amssymb}
\title{The FlySpeech Audio-Visual Speaker Diarization System for MISP Challenge 2022}
\name{
\begin{tabular}{c}
  \it Li Zhang$^1$, Huan Zhao$^1$, Yue Li$^{1}$, Bowen Pang$^1$, Yannan Wang$^{2}$,\\
  \it Hongji Wang$^{2}$, Wei Rao$^{2}$, Qing Wang$^{1}$, Lei Xie$^{1,^*}$\thanks{*Corresponding author.}
  \end{tabular}
}
\address{
  $^1$Audio, Speech and Language Processing Group (ASLP@NPU), School of Computer Science, \\
  Northwestern Polytechnical University (NPU), Xi'an, China \\
  $^2$Tencent Corporation, Shenzhen, China
}

\begin{document}
\maketitle

\begin{abstract}
This paper describes the FlySpeech speaker diarization system submitted to the second \textbf{M}ultimodal \textbf{I}nformation Based \textbf{S}peech \textbf{P}rocessing~(\textbf{MISP}) Challenge held in ICASSP 2022. We develop an end-to-end audio-visual speaker diarization~(AVSD) system, which consists of a lip encoder, a speaker encoder, and an audio-visual decoder. Specifically, to mitigate the degradation of diarization performance caused by separate training, we jointly train the speaker encoder and the audio-visual decoder. In addition, we leverage the large-data pretrained speaker extractor to initialize the speaker encoder.
Then we particularly explore the effectiveness of different frameworks of audio-visual decoder, including transformer, conformer, and cross-attention mechanism. 
In the decoding phase, we adjust the decoding frame shift to reduce boundary error and increase the stability of decoding results. 
Moreover, we introduce median filtering and secondary speaker verification as post-processing methods to smooth the predicted probabilities of speaker activities and to reduce speaker prediction errors on the single-speaker segments, respectively. 
With the above contributions, our best system achieves 10.90\% diarization error rate~(DER) on the MISP evaluation set, being ranked 3rd on the diarization track of this challenge.


\end{abstract}
\vspace{-0.65em}
\section{Introduction}
\label{sec:intro}
\vspace{-0.5em}
Speaker diarization~(SD), sometimes called active speaker detection, is the task of determining “who spoke when?” in an audio or video recording~\cite{anguera2012speaker}. Its application is indispensable in multimedia information retrieval, speaker turn analysis, and multi-speaker speech recognition. With the emergence of wide and complex application scenarios of SD, uni-modal~(audio- or visual-based) SD encounters performance bottleneck~\cite{watanabe2020chime,he2022}. Therefore, the MISP challenge 2022~\cite{chen2022first,2022misptask2} particularly set up a multi-modal SD track to promote effective integration of audio and visual information, aiming to obtain better environmental and speaker robustness in realistic applications.

In this challenge, we focus on building a robust end-to-end neural audio-visual speaker diarization (AVSD) system, which is composed of a~(visual) lip encoder, an~(audio) speaker encoder as well as an audio-visual decoder. Importantly, we take a large-data pre-trained speaker extractor to initialize the speaker encoder and train the speaker-encoder-joint-decoder system in an end-to-end manner using the challenge data.  Leveraging the pre-trained speaker extractor and the joint training strategy leads to improved system robustness and substantial SD performance gain. In the decoder module, we introduce different neural network frameworks~(transformer~\cite{dong2018speech}, conformer~\cite{gulati2020conformer}, cross-attention~\cite{tao2021someone}) to effectively fuse audio and visual representations. In addition, we adjust the decoding frame shift and make median filtering~\cite{medennikov2020target} on predicted probabilities to smooth the predicted results. Finally, we leverage the pre-trained speaker encoder to perform secondary speaker verification of single-speaker speech segments in the predicted Rich Transcription Time Masked~(RTTM). Together with the above strategies, our superior system achieves 10.90\% DER on the evaluation set at the AVSD track of MISP challenge 2022.

\vspace{-0.65em}
\section{System Description}
\begin{figure}[th]
 \captionsetup{font={footnotesize}} 
\centering
\centerline{\includegraphics[width=\columnwidth]{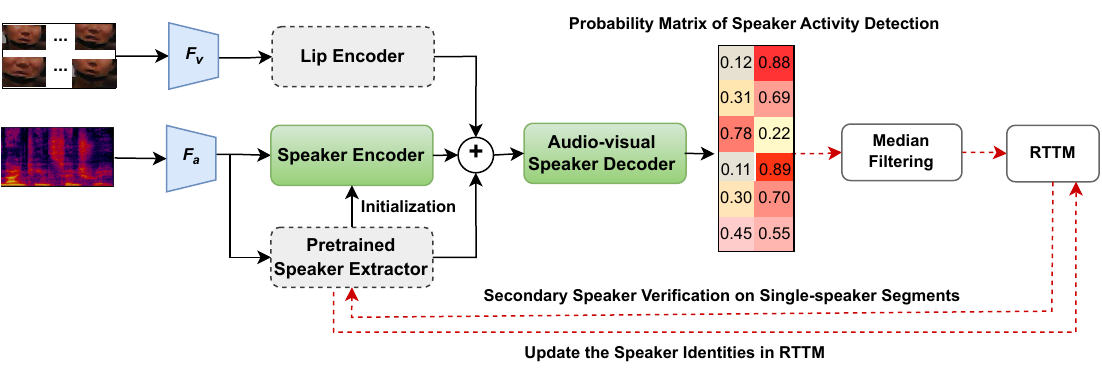}}
\caption{Overview of the submitted system. The encoders in the grey boxes are only used to extract embeddings and do not involved in optimization during joint training. The red dashed line represents post-processing. 
}
\label{fig:overview}
\vspace{-1em}%
\end{figure}

\vspace{-0.5em}
An overview of our system is illustrated in Fig~\ref{fig:overview}. It consists of visual and audio feature extractors~($F_v$ and $F_a$), uni-modal encoders~(lip encoder and speaker encoder), an audio-visual speaker decoder, and post-processing on decoding results. Specifically, the lip encoder and pretrained speaker extractor in the grey boxes are pretrained and do not participate in audio-visual decoder training. They are used to extract lip and speaker embeddings. Meanwhile, the parameters of the speaker encoder are initialized by the pretrained speaker extractor. The embeddings from lip encoder, speaker encoder, and pretrained speaker extractor are fused and fed into the audio-visual speaker decoder for speaker activation probability prediction. The audio-visual speaker decoder is to decode speaker activities of each frame into probability matrices which represent whether each speaker speaks at each frame. In the post-processing module, we first apply median filtering to smooth the predicted probabilities of speaker activity detection. Then we use secondary speaker verification (SV) on the single-speaker segments particularly to reduce speaker prediction errors in RTTM, where the pretrained speaker extractor serves as the SV model.

\textbf{Audio and Visual Feature Extraction~($F_a$ and $F_v$):} For audio feature extraction, 80-dimensional log Mel-filter bank features are extracted from input speech with 25ms window size and 10ms window shift. The visual features are the regions of interest~(ROI) of the lips generated from the position detection results provided by the challenge organizers~\cite{2022misptask2,he2022}.  

\textbf{Speaker and Lip Encoders:}  In the speaker encoder module, we exploit two kinds of neural networks -- ResNet34-SE~\cite{heo2020clova} and ECAPA-TDNN~\cite{desplanques2020ecapa}. ResNet34-SE is configured with 256 channels of squeeze-and-excitation~(SE) attention modules and the channel configurations of residual blocks are~\{32, 64, 128, 256\}. The embedding dimension of ResNet34-SE is 128. ECAPA-TDNN~\cite{desplanques2020ecapa} is configured with 1024 channels. In addition, the speaker encoder is initialized by the pretrained speaker extractor which leverages a large training dataset. But differently, the pretrained speaker extractor extracts utterance-level speaker embeddings, while the speaker encoder, joint trained with the audio-visual decoder, is used to extract frame-level speaker embeddings. The lip encoder is configured as~\cite{he2022} and it is pretrained using the lip feature and the speaker activity status provided by the RTTM .

\textbf{Audio-visual Speaker Decoder:}  For the audio-visual speaker decoder, we explore transformer~\cite{dong2018speech}, conformer~\cite{gulati2020conformer} and cross-attention~\cite{tao2021someone} respectively. The transformer decoder and conformer decoder is configured as~\cite{dong2018speech,gulati2020conformer}. For the cross-attention decoder, we first perform cross-attention between lip and utterance-level speaker embeddings extracted from the prertained speaker extractor. Then another cross-attention is further performed between the output of the first cross-attention and frame-level speaker embeddings from speaker encoder, which jointly train with audio-visual decoder. Finally, we use a set of binary classification layers to predict the speaker activation probability from the fused representation.     

\textbf{Post-processing:} In the decoding stage, we adjust the frame shift from 600 frames~\cite{he2022} to 100 frames to reduce boundary error and get stable decoding results. Then we use median filtering~\cite{medennikov2020target} to smooth the predicted probabilities of speaker activity detection. Furthermore, for the single-speaker segments produced by RTTM, we use the pretrained speaker extractor to cluster the single-speaker segments to recheck and correct the speaker identities in RTTM.  

\textbf{Datasets}: Table~\ref{tab:dataset} details the specific data for training each module in Fig.~\ref{fig:overview}. 
All datasets except MISP are open-sourced on openSLR. 
\vspace{-1em}%
\begin{table}[th]
\caption{The training data used for each module in Fig.~\ref{fig:overview}. }
\vspace{-1em}%
\resizebox{\linewidth}{!}{
\begin{tabular}{ll}
\toprule[1pt]
  Module                         & Training data (from MISP \& OpenSLR)                                                                                                      \\ \toprule[1pt]
Lip Encoder                 & MISP training set                                                                                                                   \\ 
Pretrained Speaker Extractor (ResNet34-SE) & \begin{tabular}[c]{@{}l@{}}CN-CELEB 1 \& 2, VoxCeleb 1 \& 2, MISP training set,\\ MUSAN, RIRs \end{tabular}                                                                     \\ 
Pretrained Speaker Extractor (ECAPA-TDNN) & \begin{tabular}[c]{@{}l@{}}CN-CELEB 1 \& 2, VoxCeleb 1 \& 2, Primewords, DataTang~200,\\ Aishell-1, ST-CMDS, MISP training set, MUSAN, RIRs \end{tabular} \\ 
Speaker Encoder (ResNet34-SE/ECAPA-TDNN) & \begin{tabular}[c]{@{}l@{}} MISP training set \end{tabular} \\ 
Audio-visual Speaker Decoder           & MISP training set                                                                                                    \\ 
\bottomrule[1pt]
\end{tabular}
}
\label{tab:dataset}
\vspace{-2.35em}%
\end{table}

\vspace{-0.2em}
\section{Experimental results}
\vspace{-0.5em}

Results are reported in diarization error rate~(DER) which considers false alarm~(FA), missed detection~(MISS), and speaker error~(SPKERR) simultaneously. 

In this challenge, we build four AVSD systems. 
The first three systems are \textbf{ResNet-Transformer}, \textbf{ResNet-Conformer}, and \textbf{ResNet-CrossAttention}, whose speaker encoders are ResNet34-SE and decoders are transformer, conformer, and cross-attention, respectively.
The last system is \textbf{ECAPA-Transformer}, which consists of ECAPA-TDNN speaker encoder and transformer decoder.
The experimental results of these systems and post-processing methods are illustrated in Table~\ref{tab:task2_results}. The second and the third rows in Table~\ref{tab:task2_results} show that transformer is the best choice for an audio-visual decoder compared with others. With the help of post-processing strategies~(decoder frame shift, median filtering, and secondary speaker verification), ECAPA-Transformer is the superior system, which obtains 8.53\% of DER on the development set. Finally, this system is submitted to the leaderboard, which achieves 10.90\% DER on the evaluation set. 
\begin{table}[th]
\caption{The experimental results~(\%) on the development set and evaluation set~(The last line is the result on the evaluation set).}
\resizebox{\linewidth}{!}{
\begin{tabular}{lllll}
\toprule[1pt]
\multicolumn{1}{c}{System}                              & \multicolumn{1}{c}{FA}   & \multicolumn{1}{c}{MISS}  & \multicolumn{1}{c}{SPKEER} & \multicolumn{1}{l}{DER~(Dev/Evl)}   \\ \toprule[1pt]
Baseline AVSD~\cite{he2022}                                & 4.01 & 5.86  & 3.22   & 13.09/- \\ 
 \text{ResNet-Transformer}               & 1.36 & 6.23  & 1.92   & 9.54/- \\ 
	 \text{ResNet-Conformer}     & 2.01 & 5.50  & 2.10   & 9.61/- \\ 
	 \text{ResNet-CrossAttention}               & 1.35 & 6.26  & 1.95   & 9.57/- \\ 
   \text{ECAPA-Transformer}               & 1.64 & 5.77  & 1.89  & 9.30/- \\  
      \hspace{0.5em} + Decoding Frame shift~(100)  & 1.64 & 5.62  & 1.89   & 9.15/- \\  
   \hspace{0.5em}  + Median Filtering            & 2.31 & 4.75  & 1.86   & 8.92/-  \\  
    \hspace{0.5em} + Secondary SV           & \textbf{1.95} & \textbf{4.79}  & \textbf{1.78}   &  \textbf{8.53/10.90}  \\ \bottomrule[1pt]
\end{tabular}
}
\vspace{-1.6em}%
\label{tab:task2_results}
\end{table}

\section{Conclusion}
\vspace{-0.5em}
In the AVSD track of MISP challenge 2022, we explore three strategies to improve the end-to-end AVSD system. First, we jointly train the speaker encoder and the audio-visual decoder instead of separate training. Then we explore the effectiveness of different structures for the audio-visual decoder. 
To further improve the AVSD performance, we adopt useful post-processing tricks, which are adjusting frame shift, median filtering on predicted probabilities, and secondary speaker verification on single-speaker segments. Together with the above strategies, our submission achieves 10.90\% DER on the evaluation set in the AVSD track, reaching the third place.
\vspace{-0.7em}


\footnotesize
\bibliographystyle{IEEE}
\begin{spacing}{0.98}  
\bibliography{refs}
\end{spacing}
\end{document}